\documentclass{article}
\usepackage[all]{xy}
\usepackage{amssymb,pstricks}
\begin{document}
\newtheorem{cor}{Corollary}
\newtheorem{theorem}[cor]{Theorem}
\newtheorem{prop}[cor]{Proposition}
\newtheorem{lemma}[cor]{Lemma}
\newtheorem{defi}[cor]{Definition}
\newtheorem{remark}[cor]{Remark}
\newtheorem{example}[cor]{Example}

\newcommand{\bX}{{\partial X}}
\newcommand{\cC}{\mathcal{C}}
\newcommand{\cDzu}{{{\mathcal D}_\Phi^{0,1}(X)}}
\newcommand{\cN}{{\mathcal N}}
\newcommand{\cz}{{\mathbb C}}
\newcommand{\cun}{\cC^{\infty}}
\newcommand{\ind}{\mathrm{index}}
\newcommand{\nz}{{\mathbb N}}
\newcommand{\pc}{{\Psi_\Phi(\rz^4)}}
\newcommand{\psus}{{\Psi_{\mathrm{sus}({}^\Phi\!T^*S^2)-\phi}(S^3)}}
\newcommand{\produs}{\mathrm{prod}}
\newcommand{\ptsX}{{{}^\Phi\!T^*X}}
\newcommand{\ptX}{{{}^\Phi\!TX}}
\newcommand{\rz}{{\mathbb R}}
\newcommand{\sgn}{\mathrm{sign}}
\newcommand{\Sf}{\mathrm{sf}}
\newcommand{\Spec}{\mathrm{Spec}}
\newcommand{\tf}{\tilde{f}}
\newcommand{\tn}{\tilde{\nabla}}
\newcommand{\Tr}{\operatorname{Tr}}
\newcommand{\zz}{{\mathbb Z}}
\newcommand{\vf}{\varphi}
\newcommand{\be}{\begin{equation}}
\newcommand{\ee}{\end{equation}}
\newcommand{\ba}{\begin{eqnarray}}
\newcommand{\ea}{\end{eqnarray}}
\newcommand{\no}{\nonumber\\}
\newcommand{\vect}[1]{\stackrel{\rightarrow}{#1}}
\def\openone{\leavevmode\hbox{\small1\kern-3.8pt\normalsize1}}%
\newcommand{\oo}{\openone}
\newcommand{\ab}[1]{{{\ensuremath{\mathbb{#1}}}}}
\newenvironment{proof}
{\bgroup\par\smallskip\noindent{\it Proof: }}{\rule{0.5em}{0.5em}
\egroup}
\title{Gravitational and axial anomalies for generalized Euclidean Taub-NUT 
metrics}
\author{Ion I. Cot\u aescu 
\thanks{E-mail:~~~ cota@physics.uvt.ro}\\
{\small \it West University of Timi\c soara,}\\
{\small \it V. P\^ arvan Ave. 4, RO-300223 Timi\c soara, Romania}
\and
Sergiu Moroianu 
\thanks{E-mail:~~~ moroianu@alum.mit.edu}\\
{\small \it Institutul de Matematic\u a al Academiei Rom\^ ane,}\\
{\small \it P. O. Box 1-764, RO-014700 Bucharest, Romania}
\and
Mihai Visinescu 
\thanks{E-mail:~~~ mvisin@theory.nipne.ro}\\
{\small \it Department of Theoretical Physics,}\\
{\small \it National Institute for Physics and Nuclear Engineering,}\\
{\small \it Magurele, P.O.Box MG-6, RO-077125 Bucharest, Romania}}
\date{ }
\maketitle

\begin{abstract}
The gravitational anomalies are investigated for generalized Euclidean 
Taub-NUT metrics which admit hidden symmetries analogous to the 
Runge-Lenz vector of the  Kepler-type problem. In order to evaluate 
the axial anomalies, the index of the Dirac operator for these metrics 
with the APS boundary condition is computed. The role of the Killing-Yano 
tensors is discussed for these two types of quantum anomalies.

Pacs: 04.62.+v
\end{abstract}


\section{Introduction}
In the case of gravitational interaction, a consistent perturbative 
quantization is not available, even if there exist no fermions. It is
of crucial importance in the construction of any quantum theory for 
gravitation to understand the problem of anomalies which can affect 
the conservation laws.

In the present paper we shall investigate the quantum anomalies with
regard to quadratic constants of motion in some explicit examples - 
the Euclidean Taub-Newman-Unti-Tamburino (Taub-NUT) space and its 
generalizations as it was done by Iwai and Katayama 
\cite{IK1,IK2,IK3,YM}.   

Hidden symmetries are encapsulated into St\" ackel-Killing (S-K) 
tensors. Here we consider symmetric tensors 
$k_{\mu\nu} = k_{\nu\mu}$ satisfying the S-K equation
\begin{equation}\label{SK}
k_{(\mu\nu;\lambda)}= 0
\end{equation}
where a semicolon precedes an index of covariant differentiation. For 
any geodesic with tangent (momentum vector) $p_\mu$ a S-K tensor generates a 
quadratic constant along geodesic,
\begin{equation}\label{constcl}
K = k^{\mu\nu} p_\mu p_\nu \,,\quad p_{\mu}=g_{\mu\nu}(x)\dot{x}^{\nu}\,,
\end{equation}
where $g$ is the metric tensor and the over-dot denotes the ordinary proper 
time derivative.
If we are only interested in the geodesic motion of classical scalar 
particles, then eq.(\ref{SK}) is the necessary and sufficient condition 
for the existence of a quadratic constant of motion (\ref{constcl}), as 
can be seen from the Poisson bracket of $K$ with the Hamiltonian
\begin{equation}\label{H}
H = \frac{1}{2}g^{\mu\nu} p_\mu p_\nu\,.
\end{equation} 

Passing from the classical motion to the hidden symmetries of a 
quantized system, the corresponding quantum operator analog of the 
quadratic function (\ref{constcl}) is \cite{Carter,LS}:
\begin{equation}\label{constq}
{\cal K} = D_\mu k^{\mu\nu} D_\nu
\end{equation}
where $D_\mu$ is the covariant differential operator on the 
curved manifold. Working out the commutator of 
(\ref{constq}) with the scalar Laplacian
\begin{equation}\label{Ham}
{\cal H} = D_\mu D^\mu=D_{\mu} g^{\mu\nu} D_{\nu}
\end{equation}
we get
\begin{eqnarray}
[D_\mu D^\mu, {\cal K}] =&& 2 k^{\mu\nu;\lambda} D_{(\mu}D_\nu 
D_{\lambda)} + 3 k^{(\mu\nu;\lambda)}_{~~~~~~;\lambda} D_{(\mu} 
D_{\nu)}\no
&&+ \{\frac{1}{2}g_{\lambda\sigma}( k_{(\lambda\sigma;\mu);\nu} - 
k_{(\lambda\sigma;\nu);\mu})- \frac{4}{3} 
k_\lambda^{~[\mu}R^{\nu]\lambda}\}_{;\nu} D_\mu \nonumber
\end{eqnarray}
where $R_{\mu\nu}$ is the Ricci tensor.

In the classical case, the fact that $k_{\mu\nu}$ is a S-K 
tensor satisfying eq.(\ref{SK}), assures the conservation of 
(\ref{constcl}). Concerning the hidden symmetry of the quantized 
system, the above commutator does not vanish on the strength of 
(\ref{SK}). Taking into account eq.(\ref{SK}) we get:
\begin{equation}\label{com2}
[{\cal H}, {\cal K}] = - \frac{4}{3} 
\{k_\lambda^{~[\mu}R^{\nu]\lambda}\}_{;\nu} 
D_\mu
\end{equation}
which means that in general the quantum operator ${\cal K}$ does not 
define a genuine quantum mechanical symmetry \cite{Car}. On a generic 
curved spacetime there appears a {\it gravitational quantum anomaly} 
proportional to a contraction of the S-K tensor 
$k_{\mu\nu}$ with the Ricci tensor $R_{\mu\nu}$.

In other respects, the behavior of the covariant derivatives of the spinor 
theory is also dependent on the Ricci tensor. The standard Dirac theory 
is formulated in local frames where the tetrad fields $e(x)$ and 
$\hat e(x)$  determine the form of the point-dependent Dirac 
matrices $\gamma^{\mu}$  (obeying $\{\gamma^{\mu},\gamma^{\nu}\}= 
g^{\mu\nu}{\bf 1}$) and the spin connection $\Gamma_{\mu}^{spin}$ 
of the spinor covariant derivatives which  act on the spinor field 
$\psi$ as $D_{\mu}\psi=\partial_{\mu}\psi+\Gamma_{\mu}^{spin}\psi$ \cite{NOVA}. 
Moreover, the covariant derivatives commute with $\gamma^{\mu}$ and satisfy 
\[
[D_{\mu},\,D_{\nu}]\psi=
\textstyle\frac{1}{4} 
R_{\alpha\beta\mu\nu}\gamma^\alpha\gamma^{\beta}\psi \,,
\]
where $R_{\alpha\beta\mu\nu}$ is the Riemann curvature tensor. Hereby one 
deduces the properties of the {\em standard}  Dirac 
operator ${\cal D}=\gamma^{\mu}D_{\mu}$.  
Using the identity $R_{\alpha\beta\mu\nu}\gamma^{\beta}
\gamma^{\mu}\gamma^{\nu}=-2R_{\alpha\nu}\gamma^{\nu}$   
one finds the commutation rules   
\[
\left[D_{\mu}, {\cal D}\right]\psi=-\textstyle\frac{1}{2} 
R_{\mu\nu}\gamma^{\nu}\psi
\]
which show that the squared Dirac operator, 
\[
{\cal D}^2\psi=\left({\cal H}-\textstyle{\frac{1}{4}}R\,{\bf 1}\right)\psi\,,
\]
coincides with ${\cal H}$ only when the Ricci scalar $R$ vanishes. 

Hence the conclusion is that in the Ricci-flat manifolds with  $R_{\mu\nu}=0$ 
three phenomena occur simultaneously: (I) the scalar quantum anomaly 
disappears, (II) the Dirac operator becomes the exact square root 
of the Laplace operator and (III) the covariant derivatives 
commute with ${\cal D}$. In this case the operators $D_\mu$ are 
{\em conserved} and could be taken as  momentum operators even though they do 
not commute among themselves.      

In general, when the manifold is not Ricci-flat the operators constructed from 
symmetric S-K  tensors are a source of gravitational anomalies for scalar 
fields. However, when the  S-K  tensors admit a decomposition in terms of 
antisymmetric tensors Killing-Yano (K-Y) \cite{CL} the gravitational anomaly is 
absent.

The K-Y tensors are profoundly connected with supersymmetric classical and 
quantum mechanics on curved spaces where such tensors do exist \cite{GRH}. The 
K-Y tensors play an important role in theories with spin and especially in 
the Dirac theory on curved spacetimes where they produce first-order 
differential operators, called Dirac-type operators, which 
anticommute with the standard Dirac one, $D$ \cite{CL}. When the 
K-Y tensors enter as square roots in the 
structure of several second-rank S-K tensors, they 
generate conserved quantities in pseudo-classical models for 
fermions \cite{GRH} or conserved operators in Dirac theory which 
commute with $D$.

In the pseudo-classical approach \cite{GRH} of the fermions, the absence 
of the K-Y tensors hampers the evaluation of the spin contribution 
to the conserved quantities. Passing to Dirac equation in a curved 
background, the lack of the K-Y tensors makes impossible the construction
of Dirac-type operators and hidden quantum conserved operators commuting 
with the standard Dirac one.

Having in mind that the K-Y tensors  prevent the appearance 
of gravitational anomalies for scalar field and on the other hand their 
connection with supersymmetries and  Dirac-type operators, it is 
natural to investigate their role in axial anomalies.

The importance of anomalous Ward identities in particle physics is 
widely appreciated. The anomalous divergence of the axial vector 
current in a background gravitational field was large discussed in 
the literature and directly related with the index theorem. 
In even-dimensional spaces one can define the index of a Dirac 
operator as the difference in the number of linearly independent 
zero modes with eigenvalue $+1$ and $-1$ under $\gamma_5$. The index 
is useful as a tool to investigate topological properties of the 
space, as well as in computing anomalies in quantum field theory.

In this paper we want to investigate the continuous transition from 
the case in which a hidden symmetry is described by a S-K tensor which can be 
written as a symmetrized product of K-Y tensors to the situation in which the 
K-Y tensors are absent.

In Section 2 we verify explicitly that for extended Taub-NUT 
metric the commutator (\ref{com2}) does not vanish and consequently there 
are gravitational anomalies.

In Section 3 we consider the Dirac operator on extended Taub-NUT 
spaces.
In next Sections  we compute the index of the Dirac operator for the 
generalized Taub-NUT metrics with the APS boundary condition and we 
find these metrics do not contribute to the axial 
anomaly at least for not too large deformations of the standard 
Taub-NUT metric.
This result stand in contrast with the quantum anomalies 
for scalar fields discussed  in Section 2. The result is natural 
since  the index of an operator is unchanged under continuous 
deformations of that operator. In our case this would amount to a 
continuous change in the metric and the boundary condition. 
However for larger deformations of the metric there could appear 
discontinuities in the boundary condition
and therefore the index could present jumps. Our formula for the 
index involves a computable 
number-theoretic quantity depending on the coefficients
of the metric.

In Section 6 we point out some open problems in connection with unbounded 
domains. The last Section contains some concluding remarks.

\section{Gravitational anomalies in extended Taub-NUT spaces}

The Euclidean Taub-NUT metric is involved in many modern studies in 
physics \cite{Ma,AH}. From the viewpoint of dynamical systems, the 
geodesic motion in Taub-NUT metric is known to admit a Kepler-type 
symmetry \cite{GM,GR,FH,CFH}. One can actually find the so called 
Runge-Lenz vector as a conserved vector in addition to the angular 
momentum vector. As a consequence, all the bounded trajectories are 
closed and the Poisson brackets among the conserved vectors give rise 
to the same Lie algebra as the Kepler problem, depending on the 
energy. Thus the Taub-NUT metric provides a non-trivial generalization 
of the Kepler problem.

Iwai and Katayama \cite{IK1,IK2,IK3,YM} generalized the Taub-NUT metric 
so that it still admit a Kepler-type symmetry. 

\subsection{Extended Taub-NUT spaces}

The Euclidean Taub-NUT space is a special member of the family of 
four-dimensional manifolds equipped with the isometry group 
$G_{iso}=SO(3)\otimes U(1)$. These geometries can be easily constructed 
defining the line elements in local charts with spherical coordinates 
$(r,\theta, \varphi, \chi)$; among them the first three are the usual spherical 
coordinates of the vector $\vec{x}=(x^1,x^2,x^3)$, with $|\vec{x}|=r$, 
while $\chi$ is the Kaluza-Klein extra-coordinate of this chart. The spherical 
coordinates can be associated with the Cartesian ones $(x^1,x^2,x^3,x^4)$ where 
$x^4=-\mu(\chi+\varphi)$ is defined using an arbitrary constant $\mu>0$.   

The group $SO(3)\subset G_{iso}$ has three independent one-parameter subgroups, 
$SO_i(2)$, $i=1,2,3$, each one including rotations ${\mathfrak R}_i(\phi)$, 
of angles 
$\phi\in [0,2\pi)$ around the axis $i$. With this notation any rotation 
${\mathfrak R}\in SO(3)$ in the usual Euler parametrization reads 
${\mathfrak R}(\alpha,\beta,\gamma)=
{\mathfrak R}_3(\alpha){\mathfrak R}_2(\beta){\mathfrak R}_3(\gamma)$. Moreover, 
we can 
write $\vec{x}={\mathfrak R}(\varphi,\theta,0) \vec{x}_o$ where the vector 
$\vec{x}_o=(0,0,r)$ 
is invariant under $SO_3(2)$ rotations which form its little group 
(${\mathfrak R}_3\vec{x}_o=\vec{x}_o$). The main point is to define the action 
of  
two arbitrary rotations, ${\mathfrak R}\in SO(3)$ and 
${\mathfrak R}_3\in SO_3(2)\sim U(1)$, in the spherical charts,  
$({\mathfrak R},{\mathfrak R}_3) : (r, \theta, \varphi, \chi) \to 
(r, \theta', \varphi', \chi')$, 
such that 
\begin{equation}\label{iso} 
{\mathfrak R}(\varphi', \theta', \chi')={\mathfrak R}\,{\mathfrak R}(\varphi,
\theta,\chi)
{\mathfrak R}_3^{-1}\,.
\end{equation}
Hereby it results that the Cartesian coordinates transform under 
rotations ${\mathfrak R}\in SO(3)$ as 
\begin{eqnarray}
\vec{x}&\to&\vec{x}'={\mathfrak R}\,\vec{x}\,, \label{lin}\no
x^4&\to& x^{\prime\,4}=x^4+{h}({\mathfrak R},\vec{x})\,,\label{ind} 
\nonumber
\end{eqnarray}
where the function ${h}$ is given in Ref. \cite{Ind.Repr.}. Thus, the vector 
$\vec{x}$ transforms according to an usual linear representation but 
the transformation of the fourth Cartesian coordinate is governed by a 
representation of $SO(3)$ induced by $SO_3(2)$ \cite{Ind.Repr.}.  
Furthermore, we observe that the 1-forms
\[
d\Omega(\varphi, \theta, \chi)=
{\mathfrak R}(\varphi, \theta, \chi)^{-1} d{\mathfrak R}(\varphi,\theta,\chi) 
\in so(2)
\]
transform independently on ${\mathfrak R}$ as
\[
({\mathfrak R},{\mathfrak R}_3) : d\Omega(\varphi, \theta, \chi)
\to d\Omega(\varphi', \theta', \chi')={\mathfrak R}_3 d\Omega(\varphi,\theta,
\chi)
{\mathfrak R}_3^{-1}
\]
finding that, beside the trivial quantity ${ds_1}^2=dr^2$, there are two 
types of line elements invariant under $G_{iso}$,
\begin{eqnarray}
&&{ds_2}^2=-\left< d\Omega(\varphi,\theta,\chi)^2\right>_{33}=
d\theta^2+\sin^2\theta d\varphi^2\,,\no
&&{ds_3}^2=-\frac{1}{2}Tr\left[ d\Omega(\varphi,\theta,\chi)^2\right]=
d\theta^2+\sin^2\theta d\varphi^2+(d\chi+\cos\theta d\varphi)^2\,.\label{g3}
\nonumber
\end{eqnarray}
The conclusion is that the most general form of the line element invariant 
under $G_{iso}$ is given by the linear combination 
$f_1(r){ds_1}^2+  f_2(r){ds_2}^2+  f_3(r){ds_3}^2$ involving three arbitrary 
functions of $r$, $f_1$, $f_2$ and $f_3$. 

Here it is worth pointing out that the above metrics are related to the Berger 
family of metrics on 3-spheres \cite{hitch}.  These are introduced starting with 
the Hopf 
fibration $\pi_H : S^3\to S^2$ that defines the vertical subbundle  
$V \subset  TS^3$ 
and its orthogonal complement $H \subset TS^3$ with respect to the 
standard metric $g_{S^3}$ on $S^3$.
Denoting with $g_H$ and $g_V$ the restriction of 
$g_{S^3}$ to the horizontal, respectively the vertical bundle, one finds that 
the corresponding line elements are $d{s_H}^2=\frac{1}{4}{ds_2}^2$ and 
${ds_V}^2=\frac{1}{4}({ds_3}^2-{ds_2}^2)$. For each constant $\lambda >0$ the 
Berger metric on $S^3$ is defined by the formula
\begin{equation}\label{mB}
g_{\lambda}=g_H+\lambda^2 g_V\,.
\end{equation}     

In what follows we restrict ourselves to the  {\em extended} Taub-NUT manifolds 
whose metrics are defined on ${\mathbb R}^4 - \{0\}$ by the line element
\begin{eqnarray}\label{mK}
{ds_K}^2&=&g_{\mu\nu}(x)dx^{\mu}dx^{\nu}\\
&=&f(r)(dr^2+r^2d\theta^2+r^2\sin^2\theta\, d\varphi^2)
+g(r)(d\chi+\cos\theta\, d\varphi)^2 \label{sg} 
\end{eqnarray}
where the angle variables $(\theta,\varphi,\chi)$ parametrize the sphere
$S^3$ with $ 0\leq\theta<\pi, 0\leq\varphi<2\pi, 
0\leq\chi<4\pi$, while the functions 
\begin{equation}\label{fg}
f(r) = \frac{ a + b r}{r}  ~, \quad g(r) = 
\frac{ a r + b r^2}{1 + c r + d r^2}\,.
\end{equation}
depend on the arbitrary real constants $a,\,b,\,c$ and $d$.
This line element can be written in terms of the Berger metrics as
\begin{equation}\label{mKB}
{ds_K}^2=(ar+br^2)\left(\frac{dr^2}{r^2} + 4{ds_{\lambda(r)}}^2\right)
\end{equation}
where ${ds_{\lambda(r)}}^2=(g_{\lambda(r)})_{\mu\nu}dx^{\mu}dx^{\nu}$ and
\begin{equation}\label{lamda}
\lambda(r)=\frac{1}{\sqrt{1+cr+dr^2}}\,.
\end{equation}
 
If one takes the constants 
\begin{equation}\label{standard}
c=\frac{2 b}{a}, \quad d = \frac{b^2}{a^2}
\end{equation} 
with $4m = \frac{a}{b}$, the extended Taub-NUT metric becomes the original
Euclidean Taub-NUT metric up to a constant factor. In the original 
Kaluza-Klein context the Taub-NUT parameter  $m$ is positive. 

By construction, the spaces with the metric (\ref{mK})  have the 
isometry group $G_{iso}$ and, therefore, they must have 
four Killing vectors $k^\mu_A$ labeled by an index $A=1,2,3,4$ depending 
on the parametrization of $G_{iso}$.
The usual constants of motion for particles moving in these backgrounds 
are linear in the four momentum $p_\mu$,
\begin{equation} \label{ja}
J_A = k^\mu_A \, p_\mu\,. 
\end{equation}
For a particle in extended Taub-NUT backgrounds the corresponding constants of 
motion \cite{AH,GM,GR,FH,CFH} consist of a quantity which, for negative 
mass models, can be interpreted as the ``relative electric charge"
\begin{equation} \label{q}
q = g(r) (\dot\theta + \cos\theta \dot\varphi)
\end{equation}
and the angular momentum vector
\begin{equation} \label{j}
\vec{J}=\vec{x}\times\vec{p}\,+\,q\,\frac{\vec{x}}{r} \,, \quad 
\vec{p} = f(r)\dot{\vec{x}}\,. 
\end{equation}
Notice that the form of $\vec{J}$ results from the linear representation 
(\ref{lin}) combined with the induced representation (\ref{ind}) 
\cite{Ind.Repr., NOVA}.  

The remarkable result of Iwai and Katayama is that the extended Taub-NUT 
space (\ref{sg}) still admits a conserved vector, quadratic in 
$4$-velocities, analogous to the Runge-Lenz vector of the following form
\begin{equation}\label{rl}
\vec{K} = \vec{p} \times \vec{J} + \kappa \frac{\vec{x}}{r}\,.
\end{equation}
The constant $\kappa$ involved in the Runge-Lenz vector (\ref{rl}) is
\[
\kappa = - a\,E + \frac{1}{2} c\,q^2
\]
where the conserved energy $E$ is
\begin{equation}\label{en}
E = \frac{\vec{p}^{~2}}{2 f(r)} + \frac{q^2}{2 g(r)} \,.
\end{equation}
The components $K_i=k^{\mu\nu}_i p_{\mu}p_{\nu}$ of the vector $\vec{K}$ 
(\ref{rl}) involve three S-K tensors $k^{\mu\nu}_i \,,\quad i = 1,2,3$ 
satisfying (\ref{SK}).

In other respects, the Poisson brackets between the components of $\vec{J}$ 
and $\vec{K}$ are similar to the relations known for the original 
Taub-NUT metric \cite{IK1}. In particular
\begin{equation}\label{jj} 
\{J_i, K_j\} = \epsilon_{ijk} K_k \,.
\end{equation}

\subsection{The role of the K-Y tensors}

The gravitational quantum anomaly that does not exist in Ricci-flat manifolds 
can be also absent in manifold which do not have this property if the S-K 
tensors 
have a special structure. We refer to the situation in which the S-K tensor 
$k_{\mu\nu}$ can be written as a product of K-Y tensors \cite{CL}. 

A K-Y tensor of valence $2$ is an antisymmetric tensor 
$f_{\mu\nu}$ satisfying the Killing equation
\begin{equation}\label{KY}
f_{\mu(\nu;\lambda)} = 0.
\end{equation}
The integrability condition for any solution of (\ref{KY}) is
\begin{equation}\label{intcon1}
R_{\mu\nu[\sigma}^{~~~~\tau} f_{\rho]\tau} +
R_{\sigma\rho[\mu}^{~~~~\tau} f_{\nu]\tau} = 0\,.
\end{equation}
Now contracting this integrability condition on the Riemann tensor for any 
solution of (\ref{KY}) we get
\begin{equation}\label{intcon2}
f^\rho_{~(\mu}R_{\nu)\rho} =0\,.
\end{equation} 

Let us suppose that there exist a {\it square} of the 
S-K tensor $k_{\mu\nu}$ of the form of a 
K-Y tensor $f_{\mu\nu}$ \cite{CL}:
\begin{equation}\label{kff}
k_{\mu\nu} = f_{\mu\rho} f_{\nu}^{~\rho} .
\end{equation}
In case this should happen, the S-K equation 
(\ref{SK}) is automatically satisfied and the integrability 
condition (\ref{intcon2}) becomes
\begin{equation}\label{intcon3}
k^\rho_{~[\mu}R_{\nu]\rho} =0.
\end{equation} 
It is interesting to observe that in this last equation an 
antisymmetrization rather than symmetrization is involved this 
time as compared to (\ref{intcon2}). But this relation implies  
the vanishing of the commutator (\ref{com2}) which means that the 
scalar quantum anomaly does not exist for the S-K tensors which admit a 
decomposition in terms of K-Y tensors.

In what follows we shall exemplify the role of the Killing-Yano 
tensors with regards to anomalies on the Euclidean Taub-NUT space 
and its generalizations. The (standard) Euclidean Taub-NUT space is a 
hyper-K\" ahler manifold possessing a triplet of covariantly constant K-Y 
tensors, $f^i$, $i=1,2,3$. In addition, there exist a fourth K-Y tensor, $f^Y$, 
which is not covariantly constant. The presence of this last K-Y tensor
is connected with the existence of the hidden symmetries of the 
Taub-NUT geometry which are encapsulated in three non-trivial 
S-K tensors and interpreted as the components of the so-called Runge-Lenz 
vector of geodesic motions in this space. All these S-K tensors are  
products of $f^Y$ with $f^i$ and, moreover, the manifold is Ricci-flat since 
the metric tensor can be also expressed as a product of covariantly 
constant K-Y tensors through 
$f^{i\,\mu}_{~~(\alpha} f^j_{~\beta)\mu}=-2\delta_{ij}
g_{\alpha\beta}$  \cite{NOVA}.  
Obviously, for this metric there are no gravitational anomalies for scalar 
fields. 

Concerning the generalized Taub-NUT metrics, as it was done by Iwai and 
Katayama, it was proved that the 
extensions of the Taub-NUT metric do not admit K-Y tensors, even if 
they possess S-K tensors \cite{MV,CV}. The only exception is the original 
Taub-NUT metric which possesses four K-Y tensors of valence two.

Using the S-K tensor components of the Runge-Lenz vector (\ref{rl})
we can proceed 
to the evaluation of the quantum gravitational anomaly for the extended 
Taub-NUT metric. A direct evaluation shows that the commutator 
(\ref{com2}) does not vanish. The full explicit form of this commutator 
is given in the Appendix. 
\newpage

\section{Dirac operators on generalized Taub-NUT  spaces}

Other sources of anomalies could be the operators of the Dirac theory on 
the extended Taub-NUT spaces. These have to be studied as in the case of the 
genuine Taub-NUT space \cite{NOVA} using the Cartesian coordinates instead 
of the spherical ones. For this reason we consider the Cartesian charts 
$(x^1,x^2,x^3,x^4)$ with the line elements ${d\hat s_K}^2=\frac{1}{b}{ds_K}^2$ 
which can be put in the form
\[
d\hat s_K^2=U(r) d\vec{x}\cdot d\vec{x} +V(r)(dx^4 + A_{i} dx^i)^2
\]
where we denoted      
\begin{equation}\label{(tn)}
U(r)=\frac{f(r)}{b}\,,\quad V(r)=\frac{g(r)}{b\mu^2}\,,
\end{equation}
while $A_i$ are the potentials of the Dirac magnetic monopole,  
\[
A_{1}=-\frac{\mu}{r}\frac{x^{2}}
{r+x^{3}}\,,\quad 
A_{2}=\frac{\mu}{r}\frac{x^{1}}{r+x^{3}}\,,\quad A_{3}=0\,, 
\]
giving the  
magnetic field with central symmetry
\[
\vec{B}\,={\rm rot}\, \vec{A}=\mu\frac{\vec{x}}{r^3}\,.
\]
The line element of the Cartesian charts of the original Taub-NUT space 
have to be recovered imposing the constraints (\ref{standard}) and taking 
$\mu=\frac{a}{b}=4m$ that assures the condition $U(r)V(r)=1$.

In these charts it is convenient to introduce the main orbital operators of 
the relativistic quantum mechanics in coordinate representation. We define 
the momentum operators 
\[
P_{i}=-i\left(\partial_{i}-\sqrt{UV}A_{i}\partial_{4}\right)\,,\quad 
P_{4}=-i\partial_{4}\,.
\]
that give the Laplacian operator    
\[
{\cal H}=D_{\mu}D^{\mu} = -\frac{1}{U}{\vec{P}\,}^{2}-\frac{1}{V}{P_{4}}^{2}\,,
\]
and allow one to write the angular momentum operator as 
\begin{equation}\label{(angmom)}
\vec{L}\,=\,\vec{x}\times\vec{P}-\mu\frac{\vec{x}}{r}P_{4}\,.
\end{equation} 
 
The Dirac field  must be defined in local frames 
given by tetrad fields $e(x)$ and $\hat e(x)$. Their components, which give 
us the 1-forms $\hat e^{\hat\alpha}=\hat e^{\hat\alpha}_{\mu}dx^{\mu}$ and 
the local derivatives  
$\hat\partial_{\hat\nu}=e^{\mu}_{\hat\nu}\partial_{\mu}$,  
have the usual orthonormalization properties, 
$g_{\alpha\beta} e_{\hat\mu}^{\alpha}\, e_{\hat\nu}^{\beta}=
\delta_{\hat\mu \hat\nu},\, 
g^{\alpha\beta} \hat e^{\hat\mu}_{\alpha}\, \hat e^{\hat\nu}_{\beta}=
\delta_{\hat\mu \hat\nu},\, 
\hat e^{\hat\mu}_{\alpha} e^{\alpha}_{\hat\nu}=\delta^{\hat\mu}_{\hat\nu}$. 
For the Greek indices with hats ranging from 1 to 4 the lower and upper 
positions are equivalent since the flat metric $\eta=1_{4\times 4}$ is 
Euclidean. Obviously the components of the metric tensor can be written 
as  $g_{\mu\nu}=\delta_{\hat\alpha \hat\beta}\hat e^{\hat\alpha}_{\mu}
\hat e^{\hat\beta}_{\nu}$. 
In the case of the extended Taub-NUT spaces it is convenient to choose the 
tetrad fields with the following non-vanishing components \cite{P}
\begin{eqnarray}
&&\hat e^{i}_{j}=\sqrt{U}\delta_{ij}\,, \quad
\hat e^{4}_{i}=\sqrt{V}A_{i}\,, \quad
\hat e^{4}_{4}=\sqrt{V}\,, \label{he}\no
&&e^{i}_{j}=\frac{1}{\sqrt{U}}\delta_{ij}\,,\quad
e^{4}_{i}=-\sqrt{V}A_{i}\,,\quad
e^{4}_{4}=\frac{1}{\sqrt{V}}\,.\label{e}
\nonumber
\end{eqnarray}
The commutation relations of the derivatives  
$\hat\partial_{\hat\nu}$ define the Cartan coefficients  
$C^{\,\cdot\cdot\,\hat\sigma}_{\hat\mu\hat\nu\cdot}=
e_{\hat\mu}^{\alpha} e_{\hat\nu}^{\beta}(\hat e^{\hat\sigma}_{\alpha,\beta}-
\hat e^{\hat\sigma}_{\beta,\alpha})$,
which will help us to write the spin connection in the local frames. 
   
The next step is to choose the Dirac matrices,
\[
\gamma^{i}=-i\left(
\begin{array}{cc}
0&\sigma_{i}\\
-\sigma_{i}&0
\end{array}
\right)\,,\quad
\gamma^{4}=\left(
\begin{array}{cc}
0&{\bf 1}_2\\
{\bf 1}_2&0
\end{array}
\right)\,,\quad
\gamma^{5}=\left(
\begin{array}{cc}
{\bf 1}_2&0\\
0&-{\bf 1}_2
\end{array}
\right)\,,
\]
where ${\bf 1}_2$ is the $2\times 2$ unit matrix and $\sigma_{i}$ are the Pauli 
matrices. These gamma-matrices are hermitian, satisfy   
 $\{ \gamma^{\hat\alpha},\, \gamma^{\hat\beta} \}
=2\delta_{\hat\alpha \hat\beta}$ 
and give the generators of the spinor representation 
\cite{TH} as 
\[
S^{\hat\alpha \hat\beta}=\frac{i}{4} 
\left[\gamma^{\hat\alpha},\, \gamma^{\hat\beta} \right]\,.
\]
With these ingredients we can construct the spin connection matrices,  
\begin{equation}\label{con}
\Gamma^{spin}_{\sigma}
=\frac{i}{4}\,\hat e_{\sigma}^{\hat\mu}\,S^{\hat\nu \hat\lambda}
(C_{\hat\mu \hat\nu \hat\lambda}+
C_{\hat\lambda \hat\mu \hat\nu}+C_{\hat\lambda \hat\nu \hat\mu})\,,
\end{equation}
involved in the structure of the covariant derivatives that give the Dirac 
operator 
\begin{equation}\label{DO}
{\cal D}=\gamma^{\mu}(x)D_{\mu}  
= \frac{i}{\sqrt{U}}\vec{\gamma}\cdot\vec{P}
+\frac{i}{\sqrt{V}}\gamma^{4}P_{4}
+\frac{i}{2} \frac{V}{\sqrt{U}}\gamma^{4}\vec{\Sigma}^{*}\cdot\vec{B}_{ef}
\label{(de)}
\end{equation}
where $\gamma^{\mu}(x)=\gamma^{\hat\alpha}e_{\hat\alpha}^{\mu}(x)$,   
\[
\Sigma_{i}^{*}=S_{i}+\frac{i}{2}\gamma^{4}\gamma^{i}\,,\quad
S_{i}=-\frac{1}{2}\varepsilon_{ijk}S^{jk}\,, 
\]
and $\vec{B}_{ef}\,={\rm rot}(\sqrt{UV} \vec{A})$.
By definition  the Hamiltonian operator is 
$H_D=\gamma^{5}{\cal D}$. 
Other important observables are the generators of the global symmetry, 
$P_{4}$ and the  whole angular momentum operator 
$\vec{\cal J}\,=\,\vec{L}+\vec{S}$. One can verify that its components, 
${\cal J}_{i}$, as well as $P_{4}$ are conserved in the sense that they 
commute with ${\cal D}$ and $H_D$. This means that, as was expected, the Dirac 
equation is covariant under the transformations of the universal covering group 
of $G_{iso}$.  

Using the Atiyah-Patodi-Singer index theorem 
for manifold with boundaries it was concluded that the Taub-NUT metric 
makes no contribution to the axial anomaly \cite{RS,Pope,EGH,PW}.
We specify that in the Taub-NUT spaces (where $U=V^{-1}$ and 
$\vec{B}_{ef}=\vec{B}$) the Dirac equation $H_D\psi =E\psi$ 
can be analytically solved obtaining discrete or continuous energy spectra 
that have no zero modes \cite{CVDirac}. Since this space is Ricci-flat 
the Dirac theory has the properties (II) and (III).
The four K-Y tensors of the Taub-NUT geometry  give rise to 
four Dirac operators according to the general rule that associates to any 
K-Y tensor $f$ the operator \cite{CL}
\begin{equation}\label{df}
{\cal D}_f = \gamma^{\mu} f_{\mu} ^{\nu}D_\nu
-\textstyle\frac{1}{6}f_{\mu\nu;\sigma}\gamma^{\mu}\gamma^{\nu}\gamma^{\sigma}
\end{equation}
anticommuting with ${\cal D}$ and called Dirac-type operator. 
The first three Dirac-type operators, ${\cal D}_{f^i}$, corresponding 
to the covariantly constant K-Y tensors  and ${\cal D}$ form a ${\cal N}=4$ 
superalgebra and are related among themselves through continuous 
transformations. The fourth Dirac-type operator ${\cal D}_{f^Y}$ 
is involved in the structure of the Runge-Lenz operator of the Dirac 
theory \cite{NOVA}. In any event, neither the standard Dirac operator nor the 
four Dirac-type operators produce axial anomalies.  

In the case of the extended Taub-NUT spaces the problem is more complicated 
since the spectrum of the Dirac operator (\ref{DO}) can not be analytically 
derived and, therefore, the axial anomaly must be studied using more refined 
methods. 

\section{Index formulas on compact manifolds with boundary}

Atiyah, Patodi and Singer \cite{aps1} discovered an index formula 
for first-order differential operators on manifolds with boundary 
with a non-local boundary condition.
Their index formula contains two terms, none of which is
necessarily an integer, namely a bulk term (the integral of a density in 
the interior of the manifold) and a boundary term defined in terms of the 
spectrum of the boundary Dirac operator. Endless trouble is caused in this 
theory by the requirement that the metric and the operator be of "product type"
near the boundary.

For Dirac operators on manifolds of the form
$[l_1,l_2]\times M$, where $M$ is closed, one can give another formula 
in terms of the spectral flow of the family of Dirac operators 
over the slices $\{t\}\times M, l_1\leq t\leq l_2$. 
A related formula appears in \cite{aps3} for periodic 
families. The rest of this section explains this index formula
in the case where the metric is not of product type near the boundary. 

\subsection{The spectral flow}
Let $(M,g)$ be a closed Riemannian spin manifold of odd dimension 
with a fixed spin structure,
$\Sigma$ the spinor bundle and $D$ the (self-adjoint) Dirac operator on $M$. 
Then $D$ has discrete real spectrum accumulating towards $\pm\infty$.
Moreover, the eta function
\[\eta(D,s):=\dim(\ker D)+\sum_{0\neq \lambda\in\Spec D}
|\lambda|^{-s}\sgn(\lambda)\]
is holomorphic for $\Re(s)>\dim(M)-1$ and extends meromorphically to
$\cz$. The point $s=0$ is regular \cite{aps3}, and the value 
$\eta(D,0)$ is by definition $\eta(D)$, the eta invariant of $D$.
Let $g_t$, $l_1\leq t\leq l_2$, be a smooth family of Riemannian metrics
on $M$, and $D_t$ the Dirac operator on $M$ with respect to $g_t$ 
and the fixed spin structure. Then 
\[[l_1,l_2]\ni t\mapsto f(t):=\eta(D_t)/2\in\rz\] 
is smooth modulo $\zz$, so $t\mapsto \exp(2\pi if(t))\in S^1$ is smooth. 
By the homotopy lifting property, there exists 
a smooth lift $\tf$ of $\exp(2\pi if)$
to $\rz$, the universal cover of $S^1$, uniquely determined by the condition 
$\tf(l_1)=f(l_1)$.
\[
\xymatrix{
&\rz\ar[d]^\exp\\
[l_1,l_2]\ar@{-->}[ur]^\tf\ar[r]^f&S^1
}
\]
From the definition, it is evident that $\tf(t)-f(t)\in\zz$.

\begin{defi}\label{def1}
The spectral flow of the family $\{D_t\}_{l_1\leq t\leq l_2}$
is 
\[\Sf(D_{l_1},D_{l_2}):=f(l_2)-\tf(l_2).\]
\end{defi}
This coincides with the original definition of the spectral flow for 
a path of self-adjoint Fredholm operators from \cite[Section 7]{aps3}, which
heuristically counts the net number of eigenvalues crossing $0$ in the positive 
direction. The spectral flow is clearly a path-homotopy invariant. 
Now the set of Riemannian metrics is convex inside the linear space of 
$2$-tensors. Therefore  the spectral flow of the pair $(D_{l_1},D_{l_2})$ 
is well-defined using \emph{any} $1$-parameter deformation of $g_{l_1}$ 
into $g_{l_2}$ and the associated path of Dirac operators.

\subsection{A generalized APS index formula}
Let 
\[\Pi^\pm:\cun(M,\Sigma)\to \cun(M,\Sigma)\] 
be the spectral projections associated to $D$ and the intervals
$[0,\infty)$, respectively $(-\infty,0]$. More precisely,
if $\phi_T$ is an eigenspinor of $D$ of eigenvalue $T$, then
\[
\Pi^+(\phi_T)=\left\{ \begin{array}{ll}
\phi_T& \mbox{if $T\geq 0$;}\\
0& \mbox{otherwise;} \end{array}\right.
~~~~~
\Pi^-(\phi_T)=\left\{ \begin{array}{ll}
\phi_T&\mbox{if $T\leq 0$;}\\
0& \mbox{otherwise.}\end{array}\right.
\]
If $X$ is a compact spin manifold with boundary of even dimension, then
the spinor bundles $\Sigma(\bX)$ and $\Sigma^\pm(X)_{|\bX}$ over $\bX$
are canonically identified by the Clifford action of the unit 
normal vector field.
We will need the following generalization of the Atiyah-Patodi-Singer 
index formula:
\begin{theorem}\label{gaps}
Let $(X,g^X)$ be a compact spin Riemannian manifold with boundary, and
\[\cun(X,\Sigma^+,\Pi^-):=\{\phi\in\cun(X,\Sigma^+);\Pi^-(\phi|_{\bX})=0\}.\]
Then the operator 
\[D^+:\cun(X,\Sigma^+,\Pi^-)\to\cun(X,\Sigma^-)\]
is Fredholm, and 
\[\ind(D^+)=\int_X\hat{A}(g^X)+\int_{\bX} T\hat{A}+\frac12\eta(D_{\bX})\]
where $T\hat{A}$, the transgression form of $\hat{A}$, depends on the $2$-jets
of $g^X$ at $\bX$.
\end{theorem}
\begin{proof}
The fact that $D^+$ is Fredholm is standard in the theory of elliptic boundary 
value problems, see e.g., \cite{bw}.
If the metric $g^X$ were of product type near $\partial X$, then the 
Atiyah-Patodi-Singer formula \cite{aps1} on $X_t$ would read
\begin{equation}\label{aps}
\ind(D^+)=\int_{X_t} \hat{A}(g^X)+\frac12\eta(D_{\bX})
\end{equation}
(we use the opposite orientation for $\bX$ as compared to \cite{aps1}).
In general we cannot expect such a product structure. In a collar neighborhood
defined by normal geodesic flow from $\bX$, $g^X$ takes the form 
\[g^X=dt^2+g_t\]
for $0\leq t<\epsilon$
(see \cite{bgm}), where $g_t$ is a smooth family of metrics on $\bX$.
So we first
deform smoothly the metric $g^X$ into a product metric near
$\bX$, keeping constant the metric at the boundary and outside the fixed 
collar neighborhood, using a smooth function $\psi$:

\[
h_s=dt^2+g_{\psi(s,t)},
~~~~
\psi(s,t)=\left\{
\begin{array}{ll}
t&~~\mbox{if} ~~s=0 ~~\mbox{or} ~~t>\frac{3\epsilon}{4};\\
0&~~\mbox{if} ~~t=0~;\\
0&~~\mbox{if} ~~s=1 ~~\mbox{and} ~~t\leq \frac{\epsilon}{2}.
\end{array}
\right.
\]
The index can be computed from the action of $D^+$ on Sobolev spaces:
\[D^+:H^1(X,\Sigma^+,\Pi^-)\to L^2(X,\Sigma^-).\]
The spinor bundles for different metrics are canonically identified
\cite{bgm}. 
Since by construction the vector field $\partial/\partial t$ is normal
to $\bX$ and of length $1$ for all the metrics $h_s$, it follows that
the projection $\Pi^-$, and hence also the
space $H^1(X,\Sigma^+,\Pi^-)$, do not vary with $s$.
Let $D^+_s$ be the Dirac operator corresponding to the metric $h_s$.
Then the family of bounded operators
\[D^+_s:H^1(X,\Sigma^+,\Pi^-)\to L^2(X,\Sigma^-)\]
is norm-continuous, thus the index stays 
constant during the deformation. Therefore we may compute $\ind(D^+)$
using eq.(\ref{aps})
for the metric $h_1$. 

Next we relate the $\hat{A}$ forms using the transgression form.
Consider the connection 
\[\tn:=ds\wedge \frac{\partial}{\partial s}+\nabla^s\]
on the bundle $TX$ over $[0,1]\times X$, where $\nabla^s$ is 
the Levi-Civita connection 
of the metric $h_s$. The curvature of $\tn$ decomposes in
\[\tilde{R}=R^s+ds\wedge \theta(s)\]
where $\theta(s)$ is defined by the above equality. Therefore
\[\hat{A}(\tn)=\hat{A}(\nabla^s)+ds\wedge \Theta(s)\]
and by inspection, $\Theta(s)$ depends on the $2$-jets of the
metric $g_{\psi(s,t)}$. Since $\hat{A}(\tn)$ is closed (like 
all characteristic forms), 
it follows that
\begin{equation}\label{ac}
\frac{\partial \hat{A}(\nabla^s)}{\partial s}=d\Theta(s).
\end{equation}
Define
\[T\hat{A}:=\int_0^1 \Theta(s)ds.\]
By integrating (\ref{ac})
on $[0,1]$, we get $\hat{A}(h_1)-\hat{A}(h_2)=dT\hat{A}$. 
By Stokes's formula,
\[\int_{X}\hat{A}(h_1)-\int_{X}\hat{A}(g^X)=\int_\bX T\hat{A}.\]
\end{proof}

As defined, $T\hat{A}$ depends on the function $\psi$. 
For us the important conclusion is the next corollary.
\begin{cor} \label{cors}
Let $\{g^X_l\}_{l\in\rz}$ be a smooth family of metrics on $X$, $D^+_l$
the associated family of Dirac operators on $X$, and $D_\bX^l$ the induced 
Dirac operator on $\bX$.
Then there exists a smooth function $u(l)$ such that
\[\ind(D^+_l)=u(l)+\frac12 \eta(D_\bX^l).\]
Moreover, for $l_1<l_2$,
\[\ind(D^+_{l_2})-\ind(D^+_{l_1})=\Sf(D_\bX^{l_1},D_\bX^{l_2}).\]
\end{cor}
\begin{proof}
Clearly $\hat{A}(g^X_l)$ depends smoothly on $l$. From the construction,
the transgression form is also clearly smooth in $l$ once we fix the auxiliary 
function $\psi$. We define
\[u(l):=\int_X\hat{A}(g^X_l)+\int_{\bX} T\hat{A}(g_l^X)\]
which by Theorem \ref{gaps} proves the first statement.

Using the notation from Definition \ref{def1}, 
\[
\begin{array}{ll}
\ind(D^+_{l_2})-\ind(D^+_{{l_1}})&=u({l_2})-u({l_1})+f({l_2})-f({l_1})\\
&=u({l_2})-u({l_1})+\tf({l_2})-\tf({l_1})\\
&\quad +\Sf(D_\bX^{l_1},D_\bX^{l_2}).
\end{array}
\]
Thus the smooth function $u({l_2})-u({l_1})+\tf({l_2})-\tf({l_1})$ 
is integral-valued,
and so it vanishes identically since it does at $l={l_1}$. The conclusion 
follows by setting $l=l_2$.
\end{proof}

\subsection{Index theory on a cylinder}
Let now $g^X$ be a Riemannian metric on the cylinder $X:=[l_1,l_2]\times M$. 
Endow $X$ with the product orientation, so that $\{l_1\}\times M$
is positively oriented and $\{l_2\}\times M$ is negatively oriented inside
$X$.  Let $D^+$ be the chiral Dirac operator on $X$.
For each
$t\in[l_1,l_2]$ let $g_t$ be the metric on $M$ obtained by restricting 
$g^X$ to $\{t\}\times M$. We denote by $\Sigma_t$ the spinor bundle
over $(M,g_t)$ and by $D_t$, $\Pi^\pm_t$ the Dirac operator and the 
spectral projections with respect to the metric $g_t$. 

As we mentioned above, there exist canonical identifications of the spinor 
bundle $\Sigma_t$ with $\Sigma^\pm(X)_{|\{t\}\times M}$. Consequently 
it makes sense to denote by $\phi_t$ the restriction of a positive 
spinor from $X$ to $\{t\}\times M$.

\begin{theorem}\label{th1}
Let $X=[l_1,l_2]\times M$ be a product spin manifold with a smooth metric $g^X$
as above. Set
\[\cun(X,\Sigma^+,\Pi^-):=\{\phi\in\cun(X,\Sigma^+);\Pi^+_{l_1}\phi_{l_1}=0, 
\Pi^-_{l_2}\phi_{l_2}=0\}.\]
Then 
\[\ind\left(D^+:\cun(X,\Sigma^+,\Pi^-)\to\cun(X,\Sigma^-)\right)
=\Sf(D_{l_1},D_{l_2}).\]
\end{theorem}
Note that the projection $\Pi^-_{l_2}$ equals $\Pi^+_{l_2}$ for the opposite 
orientation on $\{l_2\}\times M$, which is the one induced from $X$. 
\begin{proof}
Deform the metric $g^X$ in a neighborhood of $\{l_1\}\times M$ to a product
metric as in the proof of Theorem \ref{gaps}. As explained there, this 
deformation does not change the index. The spectral flow is also unchanged
(we noted that it depends only on the two metrics on the ends).
For $l_1<t\leq l_2$ let $X_t:=[l_1,t]\times M\subset X$.
Then Corollary \ref{cors} gives
\[
\begin{array}{ll}
\ind(D^+_t)&=u(t)+f(t)-f(l_1)\\
&=u(t)+\tf(t)-\tf(l_1)+\Sf(l_1,t).~~~\mbox{by Def.\ \ref{def1}}
\end{array}
\]
Note that both the $\hat{A}$ volume form and the transgression $T\hat{A}$, 
hence also
$u(t)$, vanish for $t$ near $l_1$ in the product region.
Thus the smooth function $u(t)+\tf(t)-\tf(l_1)$ takes values in $\zz$, 
on the other hand both $u(t)$ and $\tf(t)-\tf(l_1)$ vanish at $t=l_1$,
so $u(t)+\tf(t)-\tf(l_1)$ vanishes identically. The conclusion follows 
by setting $t=l_2$.
\end{proof}

Note that a similar statement concerning spectral boundary 
value problems appears in \cite{sss}.

\section{Harmonic spinors over Berger spheres}

Since the cohomology groups of $S^3$ vanish in dimensions $1$ and $2$,
there exists a unique spin structure on $S^3$. Let $D_\lambda$ be the
Dirac operator corresponding to the Berger metric $g_\lambda$
defined in eq.\ (\ref{mB}). Recall that 
$D_\lambda$ is essentially self-adjoint (in $L^2$)
with discrete spectrum. 

\begin{lemma}
For $\lambda<2$, $D_\lambda$ does not admit harmonic spinors.
\end{lemma}
\begin{proof}
It is easy to compute the scalar curvature of $g_\lambda$. This is 
done for instance in \cite{EGH}. Namely, $\kappa(g_\lambda)$ is constant
on $S^3$, $\kappa(g_\lambda)=(4-\lambda^2)/12$. In particular 
$\kappa(g_\lambda)$ is positive for $\lambda<2$. Lichnerowicz's formula 
proves then that $\ker D_\lambda=0$.
\end{proof}

More generally,
Hitchin \cite{hitch} computed the eigenvalues of 
$D_\lambda$. In this paper we are only interested 
in eigenvalues close to $0$. Let us recall Hitchin's result in this case.

\begin{theorem}[\cite{hitch}]\label{thhi}
Let \[\Lambda(\lambda):=\{(p,q)\in{\nz^*}^2;\lambda^2=
2\sqrt{(p-q)^2+4\lambda^2pq}\}.\]
Then
\[\dim\ker (D_\lambda)=N(\lambda):=\sum_{(p,q)\in\Lambda(\lambda)}p+q.\]
If $N(\lambda)>0$ there exists $\epsilon>0$ such that for 
$|t-\lambda|<\epsilon$, the "small" eigenvalues of $D_t$ are
given by families
\begin{equation}\label{sgab}
T(t,p,q)\,:=\frac{t}{2}-\sqrt{\frac{(p-q)^2}{t^2}+4pq},~~~
(p,q)\,\in\Lambda(\lambda)
\end{equation}
with multiplicity $p+q$.
\end{theorem}
In particular, harmonic spinors appear first for $\lambda=4$ where 
the kernel of $D_4$ is two-dimensional. Moreover, the set of those
$\lambda\in(0,\infty)$ for which $N(\lambda)\neq 0$
is discrete. For $l>0$ set 
\begin{equation}\label{eqs}
S(l):=\sum_{\lambda\leq l}N(\lambda).
\end{equation}
Of course the sum is finite for finite $l$.
\begin{cor}\label{corsf}
The spectral flow of the family $\{D_t\}_{t\in[l_1,l_2]}$ of Berger Dirac
operators equals $S(l_2)-S(l_1)$.
\end{cor}
\begin{proof}
By differentiating eq.(\ref{sgab}) we see that the function $t\to T(t,p,q)$
is strictly increasing, so the spectral flow of the family $\{D_t\}$
across $t=\lambda$ is precisely $N(\lambda)$.
\end{proof}

\section{The extended Taub-NUT metric}
Let us consider the extended Taub-NUT metric $ds^2_K$
on $\rz^4\setminus \{0\}
\simeq (0,\infty)\times S^3$
given by eq.\ (\ref{mKB}) in terms of the Berger metrics. 
We clearly need $a+br>0$ for all $r>0$ so we ask that 
$a\geq 0$, $b>0$. Also $d>0$ seems reasonable in order for the metric to 
be defined for large $r$, and even $c>-2\sqrt{d}$ so that $1+cr+dr^2>0$
for all $r>0$. 
However there seems to be no reason to ask $c\geq 0$, 
so $\lambda(r)$ may become large for certain values of $r$.

In mathematical terms, axial anomalies translate to Dirac operators 
with non-vanishing index. 
We are interested in the chiral Dirac operator on a annular piece of 
$\rz^4\setminus \{0\}$. First set
$X_{l_1,l_2}:=[l_1,l_2]\times S^3\subset \rz^4\setminus \{0\}$
with the induced extended Taub-NUT metric.
\begin{theorem}
The index of $D^+$ over $(X_{l_1,l_2},ds^2_K)$ with the APS boundary
condition is
\[\ind(D^+)=S(\lambda(l_2))-S(\lambda(l_1))\]
where the function $S$ is given by (\ref{eqs}).
\end{theorem}
\begin{proof}
By Theorem \ref{th1} the index is equal to the spectral flow of 
the pair of boundary Dirac operators. Now the metrics on the boundary spheres
are constant multiples of the Berger metrics $g_{\lambda(l_1)}$, respectively
$g_{\lambda(l_2)}$. The spectral flow of a path of conformal metrics 
(even with non-constant conformal factor) vanishes by the conformal 
invariance of the space of harmonic spinors \cite{hitch}. Thus the spectral
flow can be computed using the pair of metrics 
$g_{\lambda(l_1)}$ and $g_{\lambda(l_2)}$.
The conclusion follows from Corollary \ref{corsf}.
\end{proof}

It is a number-theoretic question to determine $S(\lambda)$ in general.
We can give however some conditions which entail the vanishing of the index.

\begin{cor}
If $c>-\frac{\sqrt{15d}}{2}$ then the extended Taub-NUT metric does 
not contribute to the axial anomaly on any annular domain (i.e., 
the index of the Dirac operator with APS boundary condition vanishes).
\end{cor}
\begin{proof}
The hypothesis implies that $\lambda(r)<4$ for all $r>0$. From the 
remark following Theorem \ref{thhi} we see that
$S(\lambda(l_1))=S(\lambda(l_2))=0$.
\end{proof}

We obtain as a particular case the vanishing of the index from \cite{EGH}.
Another case when the index vanishes is when $l_1$ and $l_2$ are either
small or large enough so that both $\lambda(l_1)$ and $\lambda(l_2)$
are less than $4$.

The singularity at the origin of the extended Taub-NUT 
metric is removable, in the sense that there exists a smooth extension 
to $\rz^4$. 

\begin{theorem}\label{thmbal}
For $l>0$ let $X_l$ be the ball $X_l:=\{r\leq l\}\subset \rz^4$, endowed
with the generalized Taub-NUT metric $ds^2_K$. Then
\[\ind(D^+)=S(\lambda(l)).\]
\end{theorem}
\begin{proof}
Deform the metric on $X_l$ smoothly into the standard metric 
$ds^2$ on the ball $X_l$. Now $ds^2=dr^2+r^2d\sigma^2$ is a warped 
product near $r=l$, so we can further deform the warping factor to be constant 
near $r=l$. Let $h_0$ be the resulting metric and $D^+_0$
its Dirac operator. The restriction of $h_0$ to $\partial X_l$ is a multiple of 
$g_1/4$, the standard metric on $S^3$. By Corollaries \ref{cors} and
\ref{corsf}, 
\[\ind(D^+)=\ind(D^+_0)+S(\lambda(l))\]
since $S(1)=0$.
We use the APS index formula (\ref{aps}) to compute $\ind(D^+_0)=0$. Indeed, 
the eta invariant of the standard sphere vanishes since the spectrum is 
symmetric around $0$,  while the $\hat{A}$ volume form of a warped product 
metric vanishes by the conformal invariance of the Pontrjagin forms.
\end{proof}

\section{Unbounded domains}

There appear two other possibilities to construct 
index problems for the metric $ds^2_K$. First we have the mixed 
APS-$L^2$ boundary condition on $[l,\infty)\times S^3$; and secondly we have
the $L^2$ index problem on $R^4$. 

The metric $ds^2_K$ is of \emph{fibered cusp} type at infinity
in the sense of \cite{mame99}.
Indeed, with the change of variables $x=1/r$ near $r=\infty$, we have
\begin{equation}\label{phi}
ds^2_K=(ax+b)\left(\frac{dx^2}{x^4}+\frac{g_H}{x^2}
+\frac{1}{d+cx+x^2}g_V\right).\end{equation}
It is impossible to present here $\Phi$-operators and the associated 
$\Phi$-calculus $\pc$; we refer the interested reader to 
\cite{mame99,boris,phiho,phind}.
The index of Dirac operators for \emph{exact} $\Phi$
metrics was computed in \cite{boris} 
under a tameness assumption on the kernel of the family of vertical 
Dirac operators. Unfortunately, (\ref{phi}) is not exact in the sense of
\cite{boris} because of the factor $d+cx+x^2$. A general but less precise 
index formula for fully elliptic $\Phi$-operators was given in 
\cite{phiho} and then improved in \cite{phind}, where the case of a 
fibration over $S^1$ is studied in detail. 

\emph{A priori} it is not at all clear if $D^+$ is Fredholm in 
$L^2(\rz^4)$, although from Theorem \ref{thmbal}, the limit
as $l\to\infty$ of the index on $X_l$ exists and equals $0$.
A general principle of Melrose's analysis of pseudodifferential algebras
asserts that an operator in such an algebra is Fredholm on appropriate Sobolev
spaces if and only if it is \emph{fully elliptic}. Before explaining
what this is, note that the corresponding statement for $\Phi$-operators
is proved in \cite{mame99}.

\subsection{Fully elliptic $\Phi$-operators}
Let $X$ denote the radial compactification of $\rz^4$.
There exists first a notion of principal symbol for $\Phi$-operators,
living on the $\Phi$-cotangent bundle, a smooth extension of 
$T\rz^4$ to $X$.

There exists additionally a "boundary symbol" map called the normal operator,
which is a star-morphism
\[\cN:\pc\to\psus\]
with values in the \emph{suspended algebra} \cite{meleta}, 
an algebra of parameter-dependent operators along the fibers 
of the Hopf fibration. A $\Phi$-operator is called fully elliptic if 
both its principal symbol and its normal operator are invertible.

\begin{theorem}
The Dirac operator on $(\rz^4,ds^2_K)$ is not fully elliptic.
\end{theorem}
\begin{proof}
The principal symbol of $D^2$ is precisely the metric $ds^2_K$, 
which extends to a Riemannian metric on $\ptsX$. This shows that
$D$ is elliptic. 

Let $u:\rz^4\to (0,\infty)$ be a function which near $x=0$ equals $ax+b$.
Define a metric $h$ on $\rz^4$ conformal to $ds^2_K$ by $h:=\frac{ds^2_K}{u}$.
Then the Dirac operators of the metrics $h$ and $ds^2_K$ are related by
\cite[Prop. 1.3]{hitch}:
\[D_h=u^{5/4}Du^{-3/4}.\]
Now notice that $u(x)=\sqrt{d}>0$ for $x=0$, and recall that the map $\cN$
is multiplicative. Thus we see that the normal operators of 
$D_h$ and $D$ are simultaneously
invertible. We focus in the rest of the proof on $D_h$. 

We want to show that $D_h$ is not fully elliptic, so we only look at the 
region $x\leq l$. Let $v(x):=\sqrt{d+cx+x^2}$, so that
\[
h=\frac{dx^2}{x^4}+\frac{g_H}{x^2}+\frac{1}{v^2(x)}g_V.
\]

Let $I,J,K$ be the three vector fields on $S^3$, viewed as the quaternion 
unit sphere, corresponding to the three distinguished complex structures.
Let $V_j, 0\leq j\leq 3$, be the following $\Phi$-vector fields
on $X$ near the boundary:
\[
V_0\,:=x^2\partial_x,~~ V_1\,:=v(x)I,~~ V_2\,:=xJ,~~ V_3\,:=xK.
\]
These vector fields form an orthonormal frame $p$ in $\ptX$, parallel
in the direction of $V_0$ (with respect to the Levi-Civita covariant 
derivative of the metric $h$). We use this frame (more precisely, one of its 
lifts $\tilde{p}$ to the Spin bundle) to trivialize
the spinor bundle. After some computations, we get
\[\cN(D_h)(0)=c^1(V_1+\sqrt{d}c^2c^3)\] 
in the above trivialization. Each integral curve $\cC$ of $V_1$
has length $2\pi/\sqrt{d}$; let $t$ be the arc-length parameter on $\cC$. 
Let $\psi$ be a spinor with 
\begin{equation} \label{ecf}
(V_1+\sqrt{d}c^2c^3)\psi=0.
\end{equation} 
We can assume that $\psi$ is a section of $\Sigma^+$, the other case is similar.
The restriction of $\psi$ to $\cC$ is given by a curve 
\[[0,2\pi/\sqrt{d})\ni t\mapsto \psi(t)\in \cz^2,\] 
where
the two factors of $\cz$ are the $\pm i$ eigenspaces of $c^2c^3$.
In other words, $\psi(t)=(\psi_+(t),\psi_-(t))$
with $c^2c^3\psi_\pm(t)=\pm i\psi_\pm(t)$.
Then eq.\ (\ref{ecf}) reduces to
\[\psi'(t)\pm i\sqrt{d}\psi_\pm(t)=0\]
and this equation does have solutions, namely 
$\psi_\pm(t)=e^{\mp it\sqrt{d}}\psi(0)$. The point is that 
the solution is periodic of period equal to the length of $\cC$. 
Equivalently, $\psi_\pm$ can be any smooth section in the complex 
line bundle over $S^2$ associated to the Hopf principal $S^1$-bundle
$S^3\to S^2$ and the $\pm 1$ representations of $S^1$ on $\cz$. 
\end{proof}

Thus our Dirac operator is not Fredholm on $L^2(\rz^4,\Sigma)$. However, 
it may still have a finite-dimensional kernel and cokernel. 
We leave open the question of determining the index in this case, 
but we conjecture it to be $0$.

The same argument shows that the Dirac operator on $[l,\infty)\times S^3$
with mixed APS-$L^2$ boundary conditions is not Fredholm either. Again, we 
leave open the question of determining its index.

\section{Concluding remarks}
There is a relationship between the absence of gravitational anomalies and
the existence of K-Y tensors.

For scalar fields, the decomposition (\ref{kff}) of S-K tensors in 
terms of K-Y tensors guarantees the absence of gravitational anomalies.
Otherwise operators constructed from symmetric tensors are in general a 
source of anomalies proportional to the Ricci tensors.

However for the axial anomaly the role of K-Y tensors is not so obvious.
The topological aspects are more important and the existence of K-Y tensors 
is not directly related to anomalies.

\subsection*{Acknowledgments}
Moroianu has been partially supported by the
RTN HPRN-CT-2002-00280
``Quantum Spaces -- Noncommutative Geometry''
funded by the European Commission, and by a CERES contract (2004); 
Cot\u aescu and Visinescu have been partially supported by the 
MEC-AEROSPATIAL Program, Romania.

\setcounter{equation}{0} \renewcommand{\theequation}
{A.\arabic{equation}}

\section*{Appendix A\\ Explicit evaluation of the gravitational anomaly for 
generalized Taub-NUT metrics}

In order to evaluate the commutator (\ref{com2}) involving the
components of the S-K tensors corresponding to the 
Runge-Lenz vector (\ref{rl}), we limit ourselves to give  
only the components of the third S-K  $k_3^{\mu\nu}$ tensor in 
spherical coordinates. Its
non vanishing components  are:
\begin{eqnarray}
&&k^{rr}_3 = -\frac{a r \cos\theta}{2(a+ b r)}\no
&&k^{r\theta}_3=k^{\theta r}_3=\frac{\sin\theta}{2}\no
&&k^{\theta\theta}_3 =\frac{(a + 2 b r) \cos\theta}{2r(a+ b r)}\no
&&k^{\varphi\varphi}_3= \frac{(a + 2 b r) \cot\theta\csc\theta}{ 2r(a+ b 
r)}\no
&&k^{\varphi\chi}_3=k^{\chi\varphi}_3=-\frac{(2a +3br + 
br\cos(2\theta)\csc^2\theta}{4r(a+br)}\no
&&k^{\chi\chi}_3= \frac{(a - adr^2 + br(2 + cr)+ 
(a + 2br))\cot^2\theta)\cos\theta}{2r(a+br)}\,.\nonumber
\end{eqnarray}

Again, just to exemplify, we write down from the 
commutator (\ref{com2}) only the function which multiplies the covariant 
derivative $D_r$:
\begin{eqnarray}
&&\frac{3r \cos\theta}{4(a+br)^3 (1 +cr +dr^2)^2}\cdot \no
&&((-a^2(c^2 +2cdr +2d(-1 +dr^2)) - 2abr(c^2 + 2cdr +2d(-1 +dr^2))+\no
&&b^2(2 +c^2r^2 +6dr^2 +2cr(2 +dr^2)))\,.\nonumber
\end{eqnarray}

Of course, as it is expected the commutator (\ref{com2}) vanishes for 
the standard Euclidean Taub-NUT metric, i.e. the constant $a,b,c,d$ are 
constrained by (\ref{standard}).

\bibliographystyle{amsplain}

\end{document}